\documentclass[aps,prb,twocolumn, superscriptaddress,amsmath,showpacs, showkeys]{revtex4}

\usepackage[latin1]{inputenc}
\usepackage{graphicx}
\usepackage{dcolumn}
\usepackage{bm}
\usepackage{setspace}

\newcommand {\dg} {\ensuremath{^{\circ}}}

\newcommand {\kvec} {\ensuremath{\mathbf{k}} vector}

\newcommand {\TN} {\ensuremath{T_{N}}}
\newcommand {\TQ} {\ensuremath{T_{Q}}}

\newcommand {\TIII} {\ensuremath{T_{\mathrm{III}}}}
\newcommand {\TIIIA} {\ensuremath{T_{\mathrm{IIIA}}}}

\newcommand {\CePrB}[2] {Ce$_{#1}$Pr$_{#2}$B$_6$}

\newcommand {\CeNdB}[2] {Ce$_{#1}$Nd$_{#2}$B$_6$}

\newcommand {\CePrBx} {\CePrB{x}{1-x}}

\newcommand {\CeNdBx} {\CeNdB{x}{1-x}}

\begin{document}

\title{Effect of Nd substitution on the magnetic order in Ce$_{x}$Nd$_{1-x}$B$_{6}$ solid solutions}




\author{J.-M. Mignot}
\author{J. Robert}
\author{G. Andr\'{e}}
\affiliation{Laboratoire L\'{e}on Brillouin, CEA-CNRS, CEA/Saclay, 91191 Gif sur Yvette (France)}

\author{M. Sera}
\author{F. Iga}
\affiliation{Department of Quantum Matter, ADSM, Hiroshima University, Higashi-Hiroshima, 739-8530 (Japan)}

\date{\today}


\begin{abstract}
Neutron powder diffraction measurements have been performed on Ce$_{x}$Nd$_{1-x}$B$_{6}$ ($x=0.5$, 0.6, 0.7, and 0.8) solid solutions to determine the type of magnetic order occurring in these compounds as a result of the interplay between magnetic dipole exchange and antiferroquadrupolar interactions. In the Ce-rich range, the sequence of two magnetic phases, with an incommensurate order [$\mathbf{k}=(1/4-\delta, 1/4-\delta, 1/2)$] forming below $T_{N}$ followed by a lock-in--type transition at lower temperature, is quite similar to that reported earlier for Ce$_{x}$Pr$_{1-x}$B$_{6}$. For $x=0.5$, on the other hand, the same antiferromagnetic order as in pure NdB$_{6}$ first occurs at $T_{N}$, then coexists with an incommensurate component below the lower transition temperature. These results are in good agreement with previous resistivity measurements and support the idea that Ce and Nd magnetic moments  in this system can be relatively decoupled.
\end{abstract}

\pacs{
75.20.Hr,		
75.25.+z,	        
75.30.Kz,	        
}


\keywords{CeB$_{6}$, NdB$_{6}$, Ce$_{x}$Nd$_{1-x}$B$_{6}$, hexaboride, neutron diffraction, magnetic phase diagram, quadrupole order, multipole interactions}

\maketitle


\section{\label{sec:intro}Introduction}

Light rare-earth hexaborides (\textit{R}B$_{6}$, \textit{R} = Ce, Pr, Nd) provide an ideal playground for studying the competition between various types of multipole interactions. CeB$_{6}$ is a dense Kondo compound, well known for developing a long-range antiferroquadrupolar (AFQ) order below a transition temperature $\TQ= 3.2$ K (phase II), then a complex magnetic dipole order below $\TIII=2.3$ K (phase III). It was first suggested from neutron diffraction experiments in an applied magnetic field,\cite{Effantin'85} then confirmed in zero field by resonant\cite{Nakao'01,Yakhou'01} and nonresonant\cite{Yakhou'01,Tanaka'04} x-ray scattering, that the order in phase II consists of a staggered arrangement, with wave vector $\mathbf{k}_{\mathrm{AFQ}}=(1/2, 1/2, 1/2)$, of $O_{xy}$-type quadrupole moments ($O_{xy}=[\sqrt{3}/2]\;[J_{x}J_{y}+J_{y}J_{x}]$) associated with the $\Gamma_{8}$ quartet ground state of Ce$^{3+}$ in the cubic crystal field (CF) of this compound. However, subsequent studies\cite{Sakai'97}  showed that neutron\cite{Effantin'85} and NMR (Ref.~\onlinecite{Takigawa'83} results can be consistently interpreted only by taking into account the extra hyperfine field produced by $T_{xyz}$ octupole moments ($T_{xyz}=[\sqrt{15}/6]\; \sum J_{x}J_{y}J_{z}$, where the summation is made over all permutations of indices). Induced octupole moments are also deemed responsible for the increase in \TQ\ observed in an applied field.\cite{Shiina'97}

In phase III, the Ce magnetic dipole moments order in a long-range, non-collinear,  so-called ``$2k$-$k'$'' magnetic magnetic structure.\cite{Effantin'85} This structure involves four different Fourier components corresponding to the commensurate wave vectors $\mathbf{k}_{C}^{(1,2)}=(1/4, \pm 1/4, 1/2)$ and $\mathbf{k}_{C}'^{(1,2)}=(1/4, \pm 1/4, 0)$, and consists of a stacking of \{001\} planes with Ce moments in adjacent planes pointing along orthogonal (in-plane) binary axes. This peculiar structure has been argued\cite{Effantin'85,Sakai'99} to reflect the competition between antiferromagnetic (AFM) dipole exchange, $O_{xy}$ AFQ, and $T_{xyz}$ antiferro-octupolar (AFO) interactions. On the other hand, the observation of similar non-collinear structures, either commensurate or incommensurate (IC), in PrB$_{6}$, which exhibits no AFQ order prior to the magnetic transition,\cite{Burlet'88} and even in GdB$_{6}$ whose ground state is a pure $S=1/2$ state with no quadrupole moment,\cite{Luca'04,Mcmorrow'04} was proposed\cite{Kuramoto'02} to arise from topological features of the Fermi surface, which is common to all hexaborides. 

In order to gain insight into the role of these different mechanisms, one can take advantage of the wide range of solid state solubility between CeB$_{6}$ and the neighboring compounds (LaB$_{6}$, PrB$_{6}$, and NdB$_{6}$). La$^{3+}$ substitution produces a dilution of the Ce sites, along with a minor expansion of the lattice, which results in the appearance of a new octupole order phase (phase IV).\cite{Mannix'05,Kusunose'05,Kuwahara'07} Pr$^{3+}$, on the other hand, has no octupole moment in his CF ground state, but can contribute to the $O_{xy}$ AFQ order.\cite{Kobayashi'01} The case of \CeNdBx\ is also interesting because pure NdB$_{6}$ orders in a simple AFM structure and shows evidence for $O_{2}^{0}$-type AFQ interactions, which can compete with the $O_{xy}$ terms.\cite{Sera'97} As expected, the magnetic phase diagrams for these compounds\cite{Yoshino'04,Kim'06} differ significantly from those of the other two series, though interesting similarities exist on the Ce-rich side. Of particular interest is the observation, from transport experiments, that the Ce--Nd exchange interaction is much weaker than the corresponding Ce--Pr interaction.\cite{Kobayashi'03}

In this paper, we present a neutron powder diffraction (NPD) study of four different \CeNdBx\ compounds on the Ce-rich side of the composition range. The results show how the magnetic properties change, with decreasing $x$, from the planar noncollinear order characteristic of Ce--Ce $O_{xy}$-type AFQ interactions, to the simple antiferromagnetic (AF) order reflecting the growing role of Nd--Nd interactions (dipole exchange and possibly $O_{2}^{0}$-type AFQ). Some of these data have been briefly presented in a previous conference report,\cite{Mignot'07} together with some single-crystal results for $x = 0.5$.

\section{\label{sec:exp}Experiments} 

\CeNdBx\ solid solutions with $x = 0.5$, 0.6, 0.7, and 0.8 were prepared at Hiroshima University using 99.5\%\ enriched $^{11}$B isotope. NPD patterns were collected at the LLB in Saclay, using the two-axis diffractometer G4-1 (800-cell position-sensitive detector) at an incident wavelength of 0.24226 nm ($x=0.5$) or 0.24266 nm ($x=0.6$, 0.7, and 0.8). A pyrolytic graphite filter was placed in the incident beam to suppress higher-order contamination. The sample powder was contained in a thin-walled vanadium cylinder, with6 mm in diameter. The data analysis was performed using the Rietveld refinement program \textsc{FullProf},\cite{fullprof'93,fullprof'01} with neutron scattering lengths and magnetic form factors taken from Refs.~\onlinecite{Sears'92} and \onlinecite{Freeman'79}, respectively. Absorption corrections were applied using an estimated absorption coefficient $\mu R = 0.36$.

\begin{table}  [b] 
\caption{\label{crysparam}Refined nuclear structure parameters of \CeNdBx.}
\begin{ruledtabular}
\begin{tabular}{c c c c c c c}
Ce concentration & $T$ (K) & $a$ (\AA) & $x_B$\\
0.5 & 4.9 & 4.1423(2) & 0.1961(3) \\
0.6 & 5.0 & 4.1443(2) & 0.1986(3) \\
0.7 & 5.0 & 4.1462(2) & 0.1986(3)\\
0.8 & 8.1 & 4.1474(2) & 0.1979(5)\\
\end{tabular}
\end{ruledtabular}
\end{table}

 \begin{figure}  
	\includegraphics [width=0.8\columnwidth, angle=0] {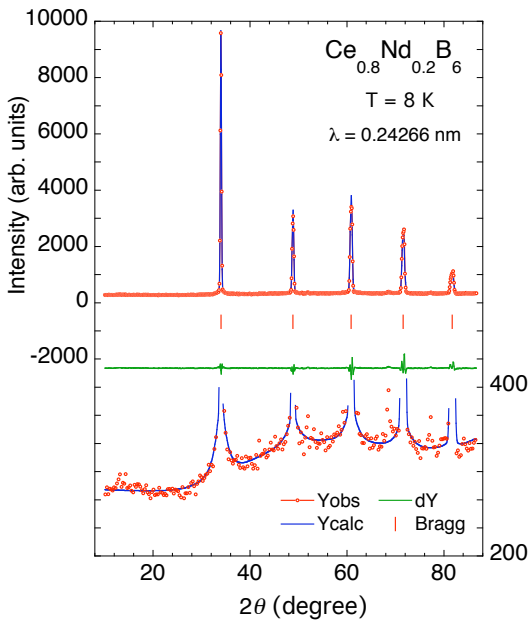}
\caption{\label{Ce0.8nuc} (Color online) Refinement of the neutron diffraction pattern of \CeNdB{0.8}{0.2}\ measured in the paramagnetic phase at $T = 8$ K (Bragg reliability factor $R_{B}=0.032$). An expanded view displaying the short-range-order contribution is plotted at the bottom of the figure (every third data point shown).}
 \end{figure}

All compounds were first measured in the paramagnetic phase above 5 K. The diffraction patterns were refined in the $Pm\bar{3}m$ crystal structure, as shown in Fig.~\ref{Ce0.8nuc} for $x = 0.8$. The resulting crystal structure parameters are listed in Table~\ref{crysparam}. The increase in the lattice parameter with increasing $x$ normally reflects the larger ionic radius of Ce$^{3+}$ in comparison with Nd$^{3+}$. The boron coherent length was also refined and the values obtained are in fair agreement with those calculated from the nominal $^{11}$B isotope content. A faint, broad, and somewhat asymmetric contribution appearing in the feet of the Bragg peaks (expanded scale in Fig.~\ref{Ce0.8nuc}), was ascribed to short-range order and fitted by an extra nuclear component with an isotropic correlation length of the order of 50 \AA. A small fraction of an unknown second phase gives rise to weak peaks above $2\theta \approx 50\dg$, but these peaks do not interfere with the magnetic signal, which is located mainly at lower angles. As in the case of \CePrBx,\cite{Mignot'08} a very weak additional intensity, with a markedly asymmetric, tow-dimensional-like, profile, is observed near $2\theta \approx 13\dg$. This signal exists for the four compositions studied, but it is more pronounced for $x=0.6$ and 0.7.


\section{\label{sec:magstrucr} Magnetic structure refinements}    

In the present measurements, the samples were first cooled down to $T_{\mathrm{min}} \approx 1.4$ K, then heated up to \TN\ by steps of typically 0.5 K. We will separately discuss the Ce-rich compounds, in which only planar structures, reminiscent of CeB$_{6}$ or PrB$_{6}$, are observed and those located near the middle of the composition range, where effects of Nd magnetism become significant.

\subsection{\label{ssec:cerich} Ce-rich compounds ($\bm{x=0.7}$, 0.8)} 

 \begin{figure*}  
	\includegraphics [width=0.80\textwidth, angle=0] {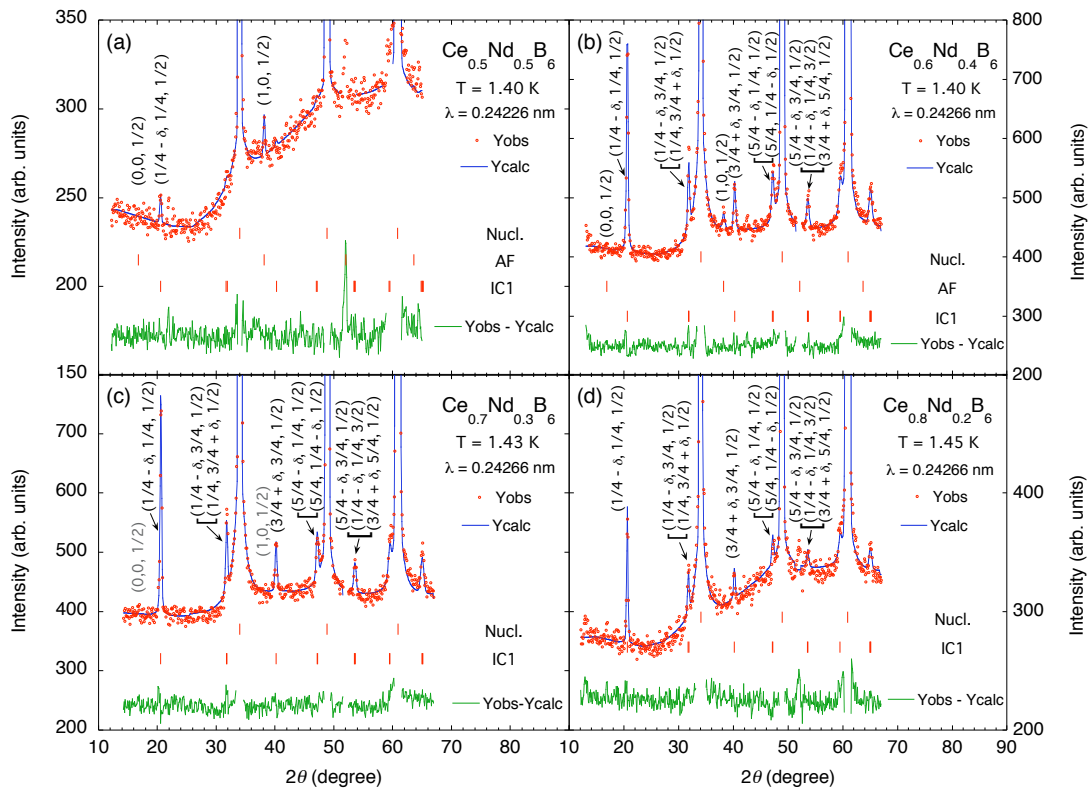}
\caption{\label{prf_T1.4} (Color online) Expanded view of the powder diffraction patterns measured at $T \approx $ 1.4 K for \CeNdBx, showing the refinement of the magnetic peaks. The indices of the lower reflections are indicated. In the difference signal, the data in the range of the nuclear peaks have been suppressed for clarity.}
\end{figure*}

At the lowest experimental temperature of 1.45 K (phase IIIA using the notation of Ref.~\onlinecite{Kobayashi'03}), the diffraction patterns exhibit one single set of magnetic satellites, which can be indexed using a magnetic \kvec\ of the same form, $(1/4-\delta, 1/4, 1/2)$, as reported in Ref.~\onlinecite{Mignot'08} for the IC1 order in \CePrBx.  The data were therefore refined, as in the latter series, assuming a planar, noncollinear, and incommensurate magnetic structure given by

\begin{eqnarray}
\mathbf{m}_{\mathrm{IC1}}(\mathbf{r}) & = & m_{\mathrm{IC1}} \left[\cos(\mathbf{k}_{\mathrm{IC1}}^{(1)}\cdot \mathbf{r})\hat{\mathbf{u}}_1 + \sin(\mathbf{k}_{\mathrm{IC1}}^{(2)}\cdot \mathbf{r})\hat{\mathbf{u}}_2 \right],
\label{eq_magstruc}
\end{eqnarray}
where
\begin{eqnarray}
\hat{\mathbf{u}}_1 & = & \frac{1}{\sqrt{2}}(1,-1,0), \nonumber \\
\hat{\mathbf{u}}_2 & = & \frac{1}{\sqrt{2}}(1,1,0)
\label{eq_unitvec}
\end{eqnarray}
are orthogonal unit vectors in the (001) plane. The superscripts (1) and (2)  denote the  \kvec\ defined above, and the one obtained by changing the sign of the second component, respectively. With this particular choice of phases in the sine and cosine functions, it can be noted that all magnetic moments have the same value $m_{\mathrm{IC1}}$.
In contrast to the complex ``$2k$-$k'$'' order observed in phase III of CeB$_{6}$.\cite{Effantin'85} no significant intensity is found here at positions corresponding to the extra  wave vector $\mathbf{k}' = (1/4-\delta, 1/4, 0)$, suggesting that the moments in adjacent (001) planes are antiparallel, as in PrB$_{6}$, rather than orthogonal as in  CeB$_{6}$. 

Considering that powder experiments cannot distinguish between single-$k$ and double-$k$ structures, refinements were performed using the simpler expression for the moment
\begin{eqnarray}
\mathbf{m}_{\mathrm{IC1}}(\mathbf{r}) & = & R_{\mathrm{IC1}} \cos(\mathbf{k}_{\mathrm{IC1}}^{(1)}\cdot \mathbf{r})\hat{\mathbf{u}}_1
\label{eq_magstruc1k}
\end{eqnarray}
as was done in Ref.~\onlinecite{Mignot'08}. The refined profiles for $x=0.8$ and 0.7 are displayed in Figs.~\ref{prf_T1.4}(d) and \ref{prf_T1.4}(c), and the parameters obtained for both concentrations are listed in Table~\ref{magmoments}. The magnetic Fourier component is seen to increase significantly between $x=0.8$ and 0.7. In the low-temperature phase, the experimental \kvec\ is very close to $(1/4,1/4,1/2)$.
\footnote{Whether the actual structure can be ensured to be IC depends on the accuracy on the $\delta$ value derived from the fit. The estimated standard deviations  given between brackets in Table~\ref{magmoments} are those calculated by \textsc{FullProf}, which are known to underestimate the probable error by about a factor of 2 because of correlated residuals.\cite{Berar'91} Therefore, we consider that the IC character of phase IIIA is likely but not fully established at this point.}
For such small  $\delta$ values, it is impossible to distinguish, between the IC1 solution used above and the IC2 solution corresponding to a \kvec\ with two incommensurate components $\mathbf{k}_{\mathrm{IC2}}=(1/4-\delta ',1/4-\delta ',1/2)$: the lower ($\mathbf{Q}=\mathbf{k}_{\mathrm{IC1,2}}$) magnetic satellite is compatible with both assumptions, provided one takes $\delta '$ equal to $\delta / 2$, and the faint splitting of the second peak ($\mathbf{Q}=\boldsymbol{\tau}_{100}-\mathbf{k}_{\mathrm{IC1,2}}$), as well as of other peaks at higher scattering angles, expected for the IC1 case cannot be ascertained within the present accuracy.

\begin{table} [t]
\caption{\label{magmoments}Refined parameters $\delta$ and $R$ and average magnetic moments $m$ in \CeNdBx\ at $T_{\mathrm{min}} \approx 1.4$ K calculated using the assumptions discussed in the text. Values for $R$ and $m$ are given in units of Bohr magnetons and for $\delta$ in reciprocal lattice units. The $B$ parameters are the magnetic Bragg reliability factors in percent.}
\begin{ruledtabular}
\begin{tabular}{c c c c c c c c c}
$x$ & $\delta$ & $R_{\mathrm{IC1}}$ & $B_{\mathrm{IC1}}$ &$R_{\mathrm{AF}}$ & $B_{\mathrm{AF}}$ &$m$\\
0.5 & 0.0060(40) & 0.55(27) & 34 & 0.48(17) & 16.6 & 0.62\\
0.6 & 0.0035(5) & 1.55(7) & 10.6 & 0.37(13) & 7.9 & 1.15\\
0.7 & 0.0027(6) & 1.46(8) & 8.1 & -- & -- & 1.03\\
0.8 & 0.0025(6) & 1.08(8) & 14.4 & -- & -- & 0.77\\
\end{tabular}
\end{ruledtabular}
\end{table}

 \begin{figure}  
	\includegraphics [width=0.55\columnwidth, angle=-90] {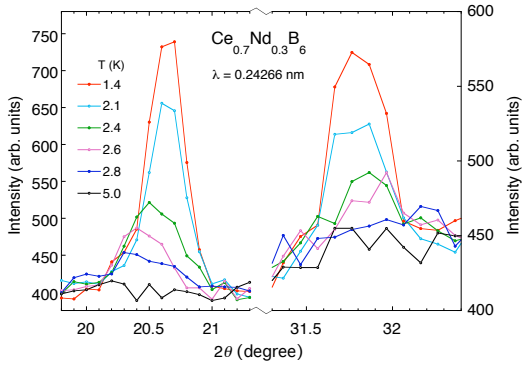}
\caption{\label{Tdpdce0.7} (Color online) Temperature dependence of the lower-angle IC magnetic satellites 000$^{\pm}$ (left) and 101$^{-}$, 011$^{-}$ (right) in \CeNdB{0.7}{0.3}.}
\end{figure}

 \begin{figure}  
	\includegraphics [width=0.65\columnwidth, angle=0] {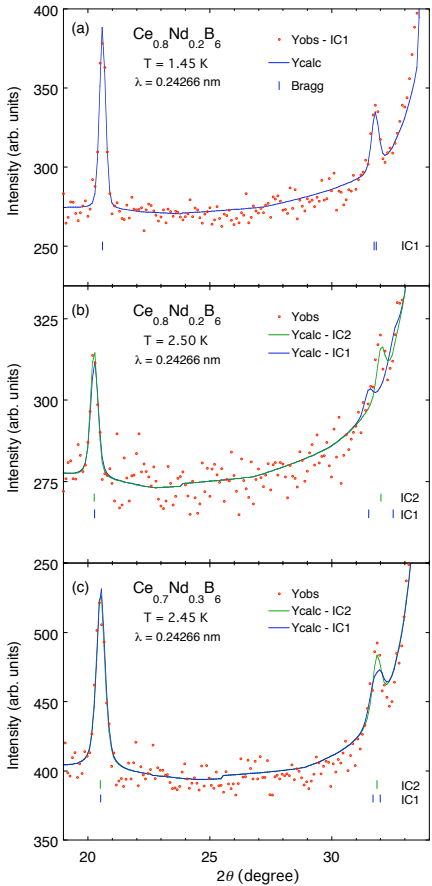}
\caption{\label{IC1/IC2} (Color online) Expanded views of the low-angle region in the diffraction patterns for \CeNdB{0.8}{0.2} at  (a) $T= 1.45$ K and  (b) 2.50 K, and  for \CeNdB{0.7}{0.3} at $T = 2.45$ K (c). In frames (b) and (c), two different refinements are shown, with or without splitting of the satellite at $2 \theta = 32\dg$, assuming IC1 or IC2 structure, respectively (see text). In frame (a), only the IC1 solution is represented but, for this set of parameters, it is indistinguishable from IC2}
\end{figure}

With increasing temperature, the same evolution is observed for both compositions. The intensities of the magnetic satellites decrease gradually over the whole temperature range but the peak positions, after remaining constant up to about 2.3 K, shift significantly above this temperature. The shift,  best evidenced by the lowest peak near $2\theta=20\dg$  (Fig.~\ref{Tdpdce0.7}), corresponds to a significant increase in the incommensurability parameter $\delta$. As a result, it becomes possible to distinguish between the IC1 and IC2 magnetic structures, mainly from the splitting, or lack thereof, of the second satellite [see Fig.~7(b) in Ref.~\onlinecite{Mignot'08}]. In both compounds [Figs.~\ref{IC1/IC2}(b) and \ref{IC1/IC2}(c)], the experimental results do favor the IC2 solution (absence of splitting). The more pronounced splitting exhibited by the calculated IC1 pattern in frame (b) reflects the larger value of $\delta$ at 2.5 K for $x = 0.8$ (see Fig.~\ref{Tdpdces} below).   Figure~\ref{IC1/IC2}(a) shows the corresponding refinement at $T=1.45$ K, in which the two solutions were undistinguishable within experimental resolution (about 0.3 \dg\ full width at half maximum near $2 \theta = 35 \dg$). The temperature dependences of the refined $R$ (magnetic Fourier components) and $\delta$ parameters are plotted in Figs.~\ref{Tdpdces}(c)--\ref{Tdpdces}(e). The present results are consistent with the transition temperatures \TN\ and \TIIIA\ derived from bulk measurements,\cite{Yoshino'04} as indicated by the arrows. It is worth noting that the variation in the magnetic amplitudes is rather smooth through the lower magnetic transition. 

\subsection{\label{ssec:midrange} Mid-range compositions ($\bm{x=0.5}$, 0.6)}    

 \begin{figure}  [b] 
	\includegraphics [width=0.75\columnwidth, angle=0] {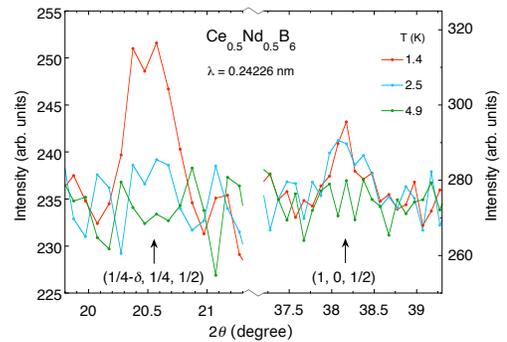}
\caption{\label{Tdpdce0.5} (Color online) Temperature dependence of the lower-angle IC 000$^{\pm}$ (left) and AF 100$^{\pm}$ (right)  magnetic satellites in \CeNdB{0.5}{0.3}.}
\end{figure}

 \begin{figure*}  
	\includegraphics [width=0.55\linewidth, angle=-90] {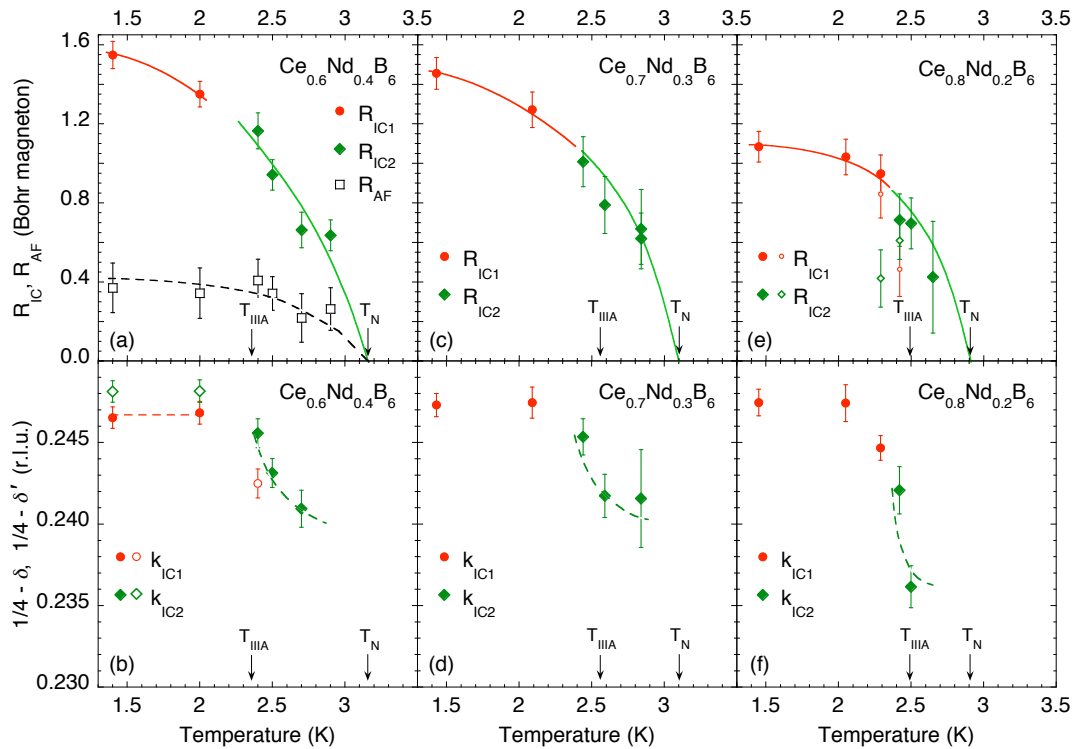}
\caption{\label{Tdpdces} (Color online) Temperature dependence of the refined magnetic Fourier components (upper frames) and the incommensurate components of the magnetic wave vectors $1/4 - \delta, 1/4 - \delta '$ (lower frames) in \CeNdBx\ for $x = 0.6$ , 0.7, and 0.8. The notations AF, IC1, and IC2 are defined in the text. The transition temperatures from Ref.~\onlinecite{Yoshino'04}, corresponding to the phase diagram in Fig.~\ref{xdpdce1.4+phdiag} below, are indicated by arrows. Error bars denote the estimated standard deviations calculated by \textsc{FullProf}, as listed in Table~\ref{magmoments}. All lines are guides for the eyes. In frame (b), open symbols represent the $\delta$ ($\delta '$) values obtained by assuming the magnetic structure to be IC1 (IC2) above (below) $T_{\mathrm{IIIA}}$. In frame (e), open symbols represent the results from an alternative refinement assuming the coexistence of IC1 and IC2 components near the transition temperature $T_{\mathrm{IIIA}}$.}
\end{figure*}

From the results of Kobayashi et al.,\cite{Kobayashi'03} the magnetic ordering pattern of \CeNdBx\ is expected to change below $x \approx 0.6$. This is indeed confirmed by the data recorded at  $T = 1.4$ K.  For $x = 0.5$, there clearly exist two distinct magnetic contributions. One is quite similar to the IC component discussed in Section \ref{ssec:cerich} but the deviation of the \kvec\ from  $(1/4,1/4,1/2)$ seems more pronounced. For the same reasons as above, we will treat this component as IC1. But there is now a second component, responsible for the extra peak near $2\theta=38.2\dg$, which denotes the same AF \kvec\ (0, 0, 1/2) previously reported for pure NdB$_{6}$.\cite{Mccarthy'80a} This structure corresponds to a doubling of the cubic unit cell along the $c$ direction, and consists of a stacking of ferromagnetic (FM) (001) planes with alternating spin orientations. The absence of a magnetic signal at the position of the (0, 0, 1/2) satellite ($2\theta=16.8\dg$) further indicates that the AF magnetic component is oriented parallel to the \kvec, i.e. normal to the FM planes, as in NdB$_{6}$. For $x=0.6$, the same two components are observed but the relative weight of the AF signal is reduced. The refined profiles are displayed in Figs.~\ref{prf_T1.4}(a) and \ref{prf_T1.4}(b), and the resulting parameters are listed in Table~\ref{magmoments}. 

At $T=2.5$ K, the diffraction pattern for $x=0.5$ (Fig.~\ref{Tdpdce0.5}) no longer contains any detectable IC satellite, and the observable magnetic signal then reduces to the AF peak at $2\theta=38.2\dg$, which remains practically unchanged. This shows that the order in the temperature region between \TIIIA\ and \TN\ is pure AF-type. A different behavior occurs for $x=0.6$ since, instead of disappearing upon heating through the lower transition at  \TIIIA, the IC component undergoes an evolution similar to that observed in the Ce-rich compounds, namely, a  \kvec\ becoming more incommensurate while changing from (presumably) IC1 to (unambiguously) IC2. Consequently, for this composition, both the AF and IC2 components appear simultaneously at \TN. Figures~\ref{Tdpdces}(a) and \ref{Tdpdces}(b) display the temperature dependences of the refined $R$ and $\delta$ parameters for this composition. As in the Ce-rich case, the data agree, within experimental accuracy, with the transition temperatures deduced from the bulk measurements. 


\section{\label{sec:discussion}Discussion}

 \begin{figure}  
	\includegraphics [width=0.8\columnwidth, angle=0] {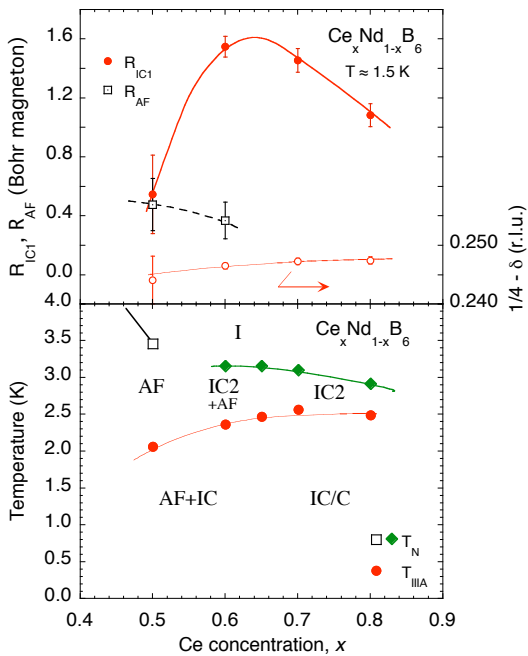}
\caption{\label{xdpdce1.4+phdiag} (Color online) (a) Concentration dependences of the IC1 and AF magnetic Fourier components (left scale) and of the IC1 propagation vector $\mathbf{k}_{\mathrm{IC1}}=(1/4-\delta,1/4,1/2)$ (open circles, right scale) derived from the refinements of the NPD patterns for \CeNdBx\ at $T\approx 1.4$ K. Lines are guides to the eye. (b) Magnetic structures observed by NPD in the different regions of the ($x$, $T$) phase diagram of  \CeNdBx (data points reproduced from Ref.~\onlinecite{Yoshino'04}).}
\end{figure}

Figure~\ref{xdpdce1.4+phdiag} summarizes the composition dependence of the magnetic ordered phases in \CeNdBx, as deduced from the present NPD measurements. The IC1 \kvec, as well as of the two types of magnetic components, is plotted in Fig.~\ref{xdpdce1.4+phdiag}(a) as a function of the Ce concentration $x$. The slight decrease in $\delta$ with increasing $x$ cannot be ascertained within experimental accuracy, though it seems plausible in view of the commensurate magnetic ground state found in pure CeB$_{6}$. The strong reduction in the IC1 component below $x=0.6$ clearly correlates with the appearance of a significant AF component. 

In Fig.~\ref{xdpdce1.4+phdiag}(b), the different regions have been mapped onto the magnetic phase diagram previously determined by Yoshino et al.\cite{Yoshino'04} One notes that, in agreement with the latter study, the composition $x=0.5$ belongs to the Nd-rich region where pure AF order, characteristic of Nd--Nd interactions, sets in at \TN. According to the present study, this AF order is not suppressed below \TIIIA\ but rather coexists with the ``IC1'' phase ascribed to Ce--Ce interactions. More surprisingly, for $x=0.6$ the AF component is still observed, although a sizable IC component now exists at all temperatures below \TN. At higher Ce concentrations, the AF component vanishes as expected. 

For the Ce-rich compounds, it is interesting to note that the observed sequence of IC phases as a function of temperature is reminiscent of the \CePrBx\ solid solutions\cite{Mignot'08} rather than of pure CeB$_{6}$. In particular, the low-temperature IC1 phase differs from ``phase III'', even disregarding the possibility of a weakly incommensurate \kvec, by the lack of extra satellites associated with the $(k_{x}, \pm 1/4, 0)$ wave vectors. This indeed implies a simple AF stacking along $z$, which could not match an AFQ order of the $O_{xy}$ moments with the wave vector (1/2, 1/2, 1/2). This result is consistent with the conclusion in Ref.~\onlinecite{Yoshino'04} that $O_{xy}$ AFQ interactions are not dominant in this phase at low fields. 

Awaji et al.\cite{Awaji'99} already pointed out that pure NdB$_{6}$, unlike  PrB$_{6}$, shows no significant effects from $O_{xy}$-type AFQ couplings, at least at low magnetic fields. The observation, in compounds with as much as 50\%\ Nd content, of an IC ordered state with a wave vector close to (1/4, 1/4, 1/2) can thus hardly be ascribed to a mere effect of AFQ interactions,\cite{Effantin'85} but rather lends support to the suggestion of Kuramoto and Kubo\cite{Kuramoto'02} that this wave vector arises from nesting properties of the Fermi surface in $R$B$_{6}$ compounds.  

Combining the different Fourier components in order to derive the magnetic structures in real space and estimate the magnetic moments at the rare-earth sites requires additional assumptions to be made. For $x=0.7$ and 0.8 this reduces to the well-known problem of defining the \kvec\ multiplicity and, in the case of \hbox{multi-$k$} structures, the phase relationship between components belonging to the same ``star''. Here we rely on arguments already given in Ref.~\onlinecite{Mignot'08}, assuming the IC1 structure to be \hbox{double-$k$}, of the type first proposed by Burlet et al.,\cite{Burlet'88} with magnetic moments of the same magnitude at all sites. In the present case, since there is only one magnetic component at $T_{\mathrm{min}}$, its value (last column in Table \ref{magmoments}) is simply equal to $R_{\mathrm{IC1}} /\sqrt{2}$. It is worth noting that the moment for $x=0.8$ is very close to that obtained in \CePrBx\ for the same composition.\cite{Mignot'08} 

For $x=0.5$ and 0.6, one has to deal with the extra complication of combining AF and IC orders. NPD data alone cannot distinguish between (i) a complex structure involving different Fourier components and (ii) a macroscopic coexistence, in separate regions of the sample, of two types of magnetic order. However, arguments in favor of the former interpretation have been derived from a careful analysis of field-induced domain repopulation effects in a single crystal.\cite{Mignot'07, Robert'up} If we therefore assume that both components contribute within the \textit{same} domain, resulting in a canting of the moments alternatively up and down along the $z$ axis, the magnetic moment at each site can be calculated, for $T \approx 1.4$ K, as the vector sum of two orthogonal components $\bm{m}_{\mathrm{IC1}}$ (within the $xy$ plane, and magnitude $R_{\mathrm{IC1}} /\sqrt{2}$) and $\bm{m}_{\mathrm{AF}}$ (along the $z$ axis, and magnitude $R_{\mathrm{AF}}$). The obtained values are listed in the last column of Table \ref{magmoments}. The increase in the moment for $x$ decreasing from 0.8 to 0.6 may be due to the larger moment of Nd, as well as to a gradual reduction in Kondo fluctuations at the Ce sites upon Nd substitution. It can be recalled, in this connection, that the single-crystal measurements\cite{Mignot'07} for $x=0.5$ have indeed revealed an additional phase, stable over a narrow temperature range just prior to the onset of the IC order, which was characterized by a steep enhancement of the AF peak intensity. That extra contribution was ascribed to a quenching of the Ce Kondo effect occurring when the Nd molecular field becomes strong enough.

The steep decrease in the ordered moment below $x=0.6$ calls for a different interpretation since Nd$^{3+}$ is normally immune from Kondo-type moment suppression. We believe that this effect actually arises from a competition between exchange and QP interactions, as was previously argued in the case of PrB$_{6}$ to explain the rather low value of the ordered Pr moment.\cite{Kobayashi'01} Here, however, the situation is more complex since two different types of QP moments, $O_{2}^{0}$ and $O_{xy}$, associated with easy magnetic axes along $z$ or in the $xy$ plane, respectively, are possibly involved. This can explain why the moment reduction sets in just at the concentration where the Nd-like AF component starts to prevail.


\section{\label{sec:conclusion}Conclusion}

The present NPD study provides an overview of magnetic ordering phenomena in the \CeNdBx\ series, from which four different regimes can be identified. At both ends of the composition range, $x < 0.4$ and $x > 0.85$ (not considered here), the compounds retain the same general behavior as in pure PrB$_{6}$ and CeB$_{6}$, respectively, and only the values of their transition temperatures and order parameters are changed.\cite{Yoshino'04} For $x$ comprised between 0.6 and 0.8, the properties are similar to those previously observed in \CePrBx, namely an incommensurate IC2 phase [$\mathbf{k}_{\mathrm{IC2}}=(1/4-\delta ',1/4-\delta ',1/2)$] forming below \TN\ and a (possibly incomplete) lock-in transition at the lower transition temperature \TIII. In analogy with the  \CePrBx\ case, it can be suggested that this transition is accompanied by a change from single-$k$ to double-$k$. The mechanism responsible for the lock-in may involve $O_{xy}$ quadrupole interactions but the lack of observation of sizable satellites associated with the $k'$ propagation vectors emphasizes the difference between the present situation and that existing in phase III of CeB$_{6}$, where the magnetic order was shown to match the (1/2,1/2,1/2) AFQ order of the $O_{xy}$ moments. Finally, concentrations between 0.4 and 0.6 represent a crossover region where AFM order, based on Nd--Nd exchange interactions, coexists with IC structures. Clear evidence was previously reported\cite{Kobayashi'03} from resistivity experiments for $x=0.4$ that Kondo scattering from Ce ions is strongly suppressed only below \TIIIA, implying that the Ce moments are taking part specifically in the low-temperature ordering process. This is consistent with the present observation that, for $x=0.5$, the IC component, characteristic of Ce--Ce interactions, also appears below the transition at \TIIIA.

Further measurements on single crystals are currently underway to check the multiplicity of the \kvec\ describing the ordered phases and to investigate the ($H$, $T$) magnetic phase diagrams in the Ce-rich composition range.

\end{document}